\documentclass[preprint]{revtex4-1}

\usepackage{graphicx}

\begin{document}

\title[Enhancing cell wall permeability by microstreaming flow]{Enhancing microalgal cell wall permeability \\ by microbubble streaming flow}

\author{Angelo Pommella}
    \altaffiliation[Now at: ]{INSA-Lyon, CNRS MATEIS UMR5510, Université de Lyon, Villeurbanne F-69621, France.}

\author{Irina Harun}
    \altaffiliation[Now at: ]{Department of Environment, Universiti Putra Malaysia, Serdang 43400, Selangor, Malaysia.}

\author{Klaus Hellgardt}

\author{Valeria Garbin}
	\altaffiliation[Now at: ]{Department of Chemical Engineering, Delft University of Technology, 2629 HZ Delft, Netherlands}
		\email{v.garbin@tudelft.nl}

\affiliation{Department of Chemical Engineering, Imperial College London, London SW7 2AZ, United Kingdom}

\date{10 December 2021}

\begin{abstract}
We demonstrate that the permeability of the cell wall of microalga \textit{C. reinhardtii} can be enhanced in controlled microstreaming flow conditions using single bubbles adherent to a wall and driven by ultrasound near their resonance frequency. We find that microstreaming flow is effective at acoustic pressures as low as tens of kPa, at least one order of magnitude lower than those used in bulk ultrasonication. We quantify the increase in number of fluorescent cells for different acoustic pressure amplitudes and ultrasound exposure times. We interpret the increase in dye uptake after microstreaming flow as an indication of enhanced permeability of the cell wall. We perform microscopic visualizations to verify the occurrence of non-spherical shape oscillations of the bubbles, and to identify the optimal conditions for dye uptake. In our experimental conditions, with acoustic pressures in the range 20-40~kPa and frequencies in the range 25-70~kHz, we obtained spherical oscillations of bare microbubbles and non-spherical shape oscillations in microbubbles with algae attached to their surface. Only the latter were able to generate a non-linear microstreaming flow with sufficiently large viscous stresses to induce algae permeabilization. The results demonstrate that controlled microbubble streaming flow can enhance cell wall permeability in energy-efficient conditions. 
\end{abstract}

\maketitle

The transport of molecules across the cell wall of microorganisms is central to biotechnology and bioseparations: in biotechnology, genetic material is delivered to microorganisms  \cite{walker1998yeast} while in bioseparations, intracellular products are extracted from microorganisms \cite{ladisch2001bioseparations}. The cell wall is an extracellular structure characteristic of both microorganisms and plant cells, a few nm to hundreds of nm thick, with a complex architecture \cite{Goodenough1985} based on polysaccharides, glycoproteins and other components. Even though they have very different biomolecular compositions and structures, cell walls are typically a mechanically strong barrier from the external environment, with tensile strengths of up to 1~GPa \cite{Carpita1985}. Methods to enhance the permeability of the cell wall of microorganisms or plant cells include among others electroporation \cite{CHASSY1988303}, hydrodynamic cavitation \cite{mevada2019} and ultrasonic treatment \cite{QIN2013247,Azencott2007,WU2016210}. 

Ultrasound-based methods to enhance the permeability of cell walls are considered to be attractive because they can be operated continuously and they can be scaled up. They share with other mechanical methods the advantage that they do not involve the use of aggressive solvents, but also the limitation of being energy intensive. Estimates of the specific energy input for cell wall disruption based on single-cell measurements \cite{Lee2013} indicate that the efficiency of mechanical methods is currently very low. Intriguingly, it has recently been shown that the efficiency of ultrasonic treatment can be increased by pre-mixing microbubbles with the cell suspension \cite{Krehbiel2014}. This is not surprising, given that in ultrasonication a significant amount of energy is dissipated to create cavitation bubbles, and further energy is lost by viscous dissipation and heating of the liquid.Controlled cell lysis experiments on microorganisms in microfluidics \cite{Ohl2012} used acoustic pressure amplitudes as high as 1~MPa, which cause bubble nucleation followed by violent cavitation dynamics \cite{Plesset1977}. This bubble dynamics regime generates large viscous stresses, on the order of $10^5$~Pa, causing complete disruption of cell walls \cite{Ohl2012} and removal of bacterial biofilms \cite{Vyas2020}. 

In this Letter we demonstrate that microbubble streaming flow can be used to enhance the uptake of a lipid-soluble fluorescent dye in unicellular microalga \textit{Chlamydomonas reinhardtii} in energy-efficient conditions. Microbubble streaming flow is a much less violent regime of ultrasound-driven bubble dynamics, which can be obtained using low acoustic pressure amplitudes on the order of 10-100~kPa, and bubbles positioned in controlled geometries. This flow regime offers tuneable flow conditions at the microscale \cite{Fauconnier2020}, microparticle manipulation and sorting \cite{Hilgenfeldt2012, Hilgenfeldt2016} and is increasingly finding applications in advanced cell sorting assays  \cite{feng2019on-chip, lu2019parallel,jiang2021rapid}. Microbubble streaming flow has been shown to be sufficient to rupture giant unilamellar lipid vesicles  \cite{Marmottant2003,Pommella2015}, and enhance the permeability of animal cell membranes  \cite{Pereno2018}. Because these systems only possess a 4-nm thick lipid membrane, and no cell wall, they are easily deformed \cite{BarthesBiesel2016} and ruptured in a flow. Whether the mild flow conditions of microstreaming are sufficient to permeabilize cell walls, which are thicker and stiffer, has not yet been proven.  

\begin{figure}[ht]
\begin{center}
\includegraphics[width= 0.85\textwidth]{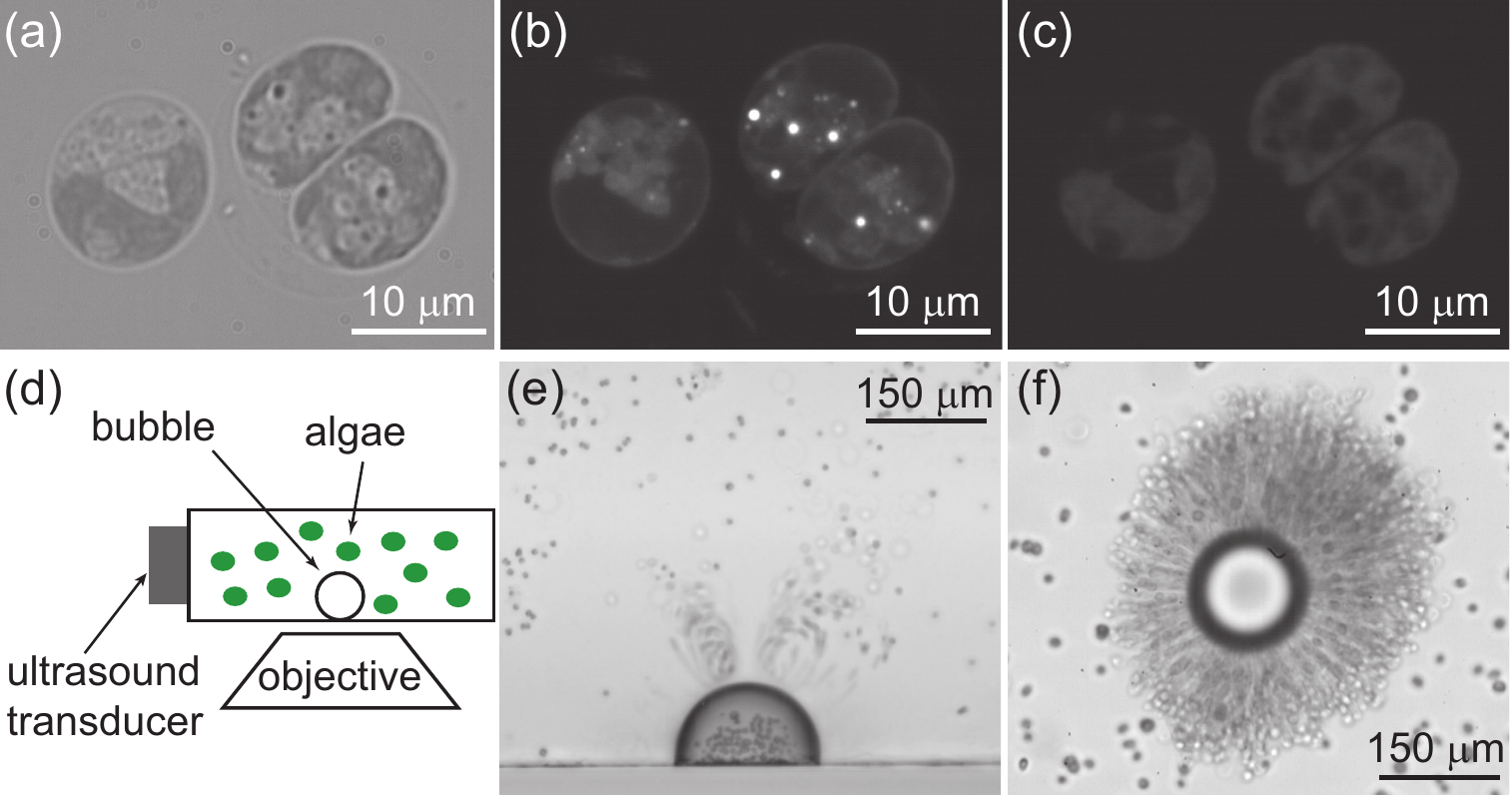}
\caption{(a-c) High-magnification optical micrographs of three algal cells acquired in (a) bright field, (b) green fluorescence channel and (c) red fluorescence channel, respectively. In the green channel, BODIPY shows the presence of lipid globules recognisable as white spots inside the cells. In the red channel, the cell autofluorescence from the chlorophyll is observed. (d) Schematic of the experimental setup for microstreaming flow experiments (not to scale). (e) Side view and (f) top view of algal cell suspension subjected to acoustic microstreaming flow induced by ultrasound-driven bubble oscillations.} \label{fig:method}
\end{center}
\end{figure}

We test the proposed method on unicellular eukaryotic green alga \textit{Chlamydomonas reinhardtii}, a widely studied model organism for biofluid dynamics  \cite{Goldstein2015}. The \textit{C. reinhardtii} cell is about 10~$\mu$m in diameter. It has a chloroplast for photosynthesis; two anterior flagella for motility; and a cell wall with seven principal layers, made primarily of glycoproteins \cite{Goodenough1985}. Wild-type \textit{C. reinhardtii} strain cc-124 cells (kindly provided by Prof. P. Nixon, Department of Life Sciences, Imperial College London) were cultured in nitrogen starvation as described in \textbf{Supporting Information}, so as to increase the lipid content. To detect cell wall permeabilization of microalgae induced by acoustic microstreaming flow, we then use changes in uptake of a fluorescent dye, BODIPY, that accumulates into the lipid bodies. The protocol for fluorescent cell staining is described in \textbf{Supporting Information}. Figures~\ref{fig:method}(a-c) show three microscopy images of a group of \textit{C. reinhardtii} algal cells after initial fluorescent staining with BODIPY and before the microstreaming flow is applied. The bright field image in Figure~\ref{fig:method}(a) shows the morphology of the algal cells. The fluorescence image (green channel) in Figure~\ref{fig:method}(b) shows that the intracellular lipid globules have become fluorescent, due to partitioning in the oil phase of BODIPY during incubation, showing that the dye can cross the cell wall \cite{Rumin2015}. The fluorescence microscopy image in Figure~\ref{fig:method}(c) is taken in the red channel and shows the typical autofluorescence of chlorophyll in the chloroplasts.

A schematic of the setup for microstreaming flow experiments \cite{Pommella2015} is shown in Figure~\ref{fig:method}(d). After initial staining with BODIPY, the algal cell suspension was transferred into a quartz cuvette. Air microbubbles of 30-120~$\mu$m in radius were formed with a syringe. Adhesion of the bubbles to the cuvette walls was promoted by making the quartz surface hydrophobic by surface treatment \cite{Pommella2015}. Care was taken to inject only a few bubbles, so that the distance between neighboring bubbles was sufficiently large, \textit{i.e.}, in excess of 10 bubble diameters, to prevent coupling of their flow fields. The bubbles were activated by the ultrasound field generated by a piezoelectric transducer (Physik Instrumente, model P-111.05) glued to the bottom of the cuvette. The transducer was driven by the signal produced by a waveform generator (Agilent) and amplified by a linear radio-frequency power amplifier (T\&C Power Conversion). In each experiment, an isolated bubble was centered in the field of view, and the frequency of the ultrasound wave was selected so as to drive the bubble near its resonance frequency, $f_{0}$, for which the amplitude of oscillations is a maximum. The resonance frequency was estimated taking into account the correction for a bubble adhered to a solid surface \cite{Blue1967}:
\begin{equation}
f_{0}=\frac{1}{2\pi R_{0}}{\sqrt{\frac{3\kappa p_0}{\rho \left ( 1+\cos \alpha  \right )}}},
\label{eq:Bluef}
\end{equation}
where $R_{0}$ is the equilibrium radius of the bubble, $\alpha$ the contact angle (measured in the aqueous phase), $\kappa$ is the polytropic exponent, $\rho$ is the liquid density, and $p_0$ is the atmospheric pressure. We observe that during activation by ultrasound the bubbles sometime decrease in radius by up to 5\%, due to gas diffusion, but their resonance frequency remains sufficiently close to  the driving frequency that they continue to oscillate and generate a microstreaming flow. Since the gas compression can be approximated as being isothermal for the bubble radii and frequencies used \cite{Plesset1977}, we assumed $\kappa \approx 1$. An average contact angle $\alpha = 46^{\circ} \pm 2^{\circ}$ was measured in experiments with different bubbles ($n=13$). The frequency of the ultrasound wave, \textit{f}, was selected in the range $0.9 f_0 < f < 1.6 f_0$ (25-70 kHz), the acoustic pressure amplitude $p_{\mathrm{A}}$ was varied in the range 20-40 kPa, and the ultrasound exposure time $t$, during which the ultrasound was applied, was varied in the range 1-60 s.  The acoustic pressure $p_A$ was calibrated as a function of ultrasound frequency using a hydrophone (RP 33 s, RP Acoustics). The pressure was measured in three different locations in the glass cuvette (bottom, middle, and top) filled with water and mounted on a holder. The average of the three measurements gave a maximum standard deviation of 10$\%$.  The linearity of the output of the power amplifier was confirmed for several values of the gain; pressure values in the experiments were then extrapolated from the linear dependence of the pressure on the gain. 

The characteristic microstreaming flow profile, with closed streamlines around the bubble, is shown in Figure~\ref{fig:method}(e-f). We perform experiments both in the linear and in the non-linear microstreaming flow regimes. We have previously shown \cite{Pommella2015} that in this experimental setup the linear microstreaming flow is obtained for acoustic pressures $\leq$ 15 kPa.  Algal cells entrained in the microstreaming flow were observed through an inverted microscope (IX71, Olympus) in both side and top view as shown in Figures~\ref{fig:method}(e) and~\ref{fig:method}(f) respectively. Microscopy images and videos at a frame rate of 10 frames per second were acquired using a camera (DCC3240M, Thorlabs) both in bright field and green fluorescence channel before and after the microstreaming flow was applied. After each experiment the sample was discarded, the cuvette was filled with fresh cell suspension, and a new bubble was generated. 

\begin{figure}[hb]
\begin{center}
\includegraphics[width= 14cm]{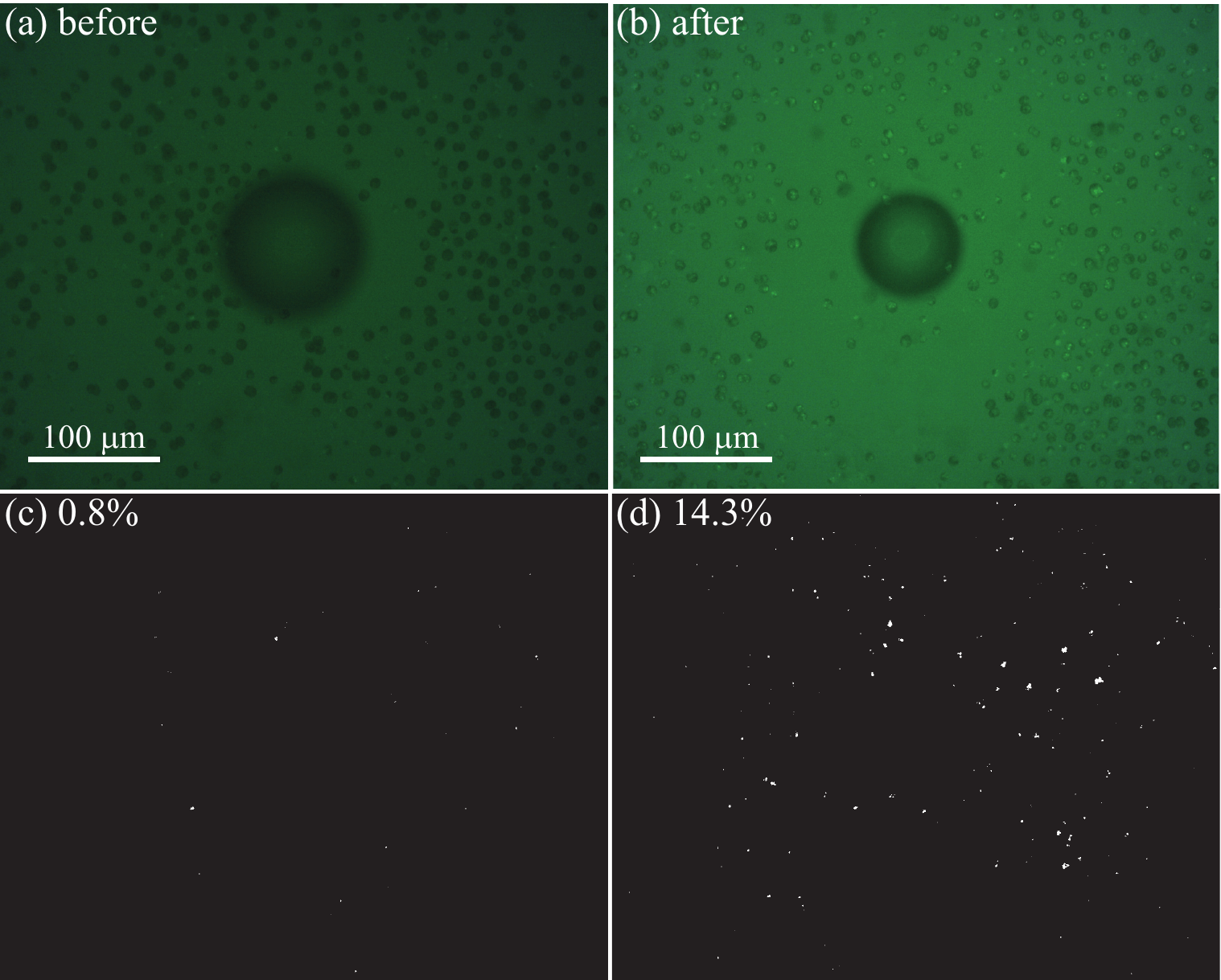}
\caption{(a, b) Images of a microbubble surrounded by algal cells before and after the acoustic microstreaming flow respectively. The images were acquired in the green channel where cells show fluorescence due to BODIPY dye. The microbubble has an initial radius of 53 $\mu$m and ultrasound at frequency of 53 kHz and pressure and 22.3 kPa was applied for 20 seconds. At the end of the ultrasound pulse, the bubble radius has decreased due to gas diffusion. (c, d) Masks of the images a e b respectively. The threshold gray value was chosen to highlight the fluorescent algal cells (white spots) discarding the fluorescent background. The percentage values indicate the amount of fluorescent cells present before (c) and after (d) the flow.} \label{fig:fluorescence}
\end{center}
\end{figure}

In Figure~\ref{fig:fluorescence} we demonstrate that acoustic  microstreaming  flow  induced  by  ultrasound-driven microbubbles can enhance the uptake of the flourescent dye BODIPY in \textit{Chlamydomonas reinhardtii} algal cells. Figure~\ref{fig:fluorescence}(a) shows the top view of a microbubble (at the centre of the image) surrounded by algal cells in the fluorescent green channel before the microstreaming flow. In Figure~\ref{fig:fluorescence}(b) we show the same microbubble after stopping the flow. The acoustic microstreaming flow was generated for 20 seconds by applied ultrasound waves at a frequency of 53 kHz and pressure of 22.3 kPa. By comparing the two images before and after the flow, an increase of the number of fluorescent algal cells is visible in Figure~\ref{fig:fluorescence}(b). This result is highlighted in the masks of Figure~\ref{fig:fluorescence}(c) and Figure~\ref{fig:fluorescence}(d) where the fluorescent algal cells of Figure~\ref{fig:fluorescence}(a) and Figure~\ref{fig:fluorescence}(b) respectively are highlighted as white spots on a dark background. After the flow stops, processed cells can still be seen swimming.  We quantified the increment of fluorescent cells by determining the percentage of fluorescent cells before and after the flow. The percentage increases from 0.8\% before the flow, to 14.3\% after the flow. The increased uptake suggests that the microstreaming flow enhances the cell wall permeability of algal cells. The difference in fluorescence background between Figures~\ref{fig:fluorescence}(a) and (b) is due to the different camera settings used to acquire the images; we describe in \textbf{Supporting Information} the image analysis protocol used to quantify the percentage of fluorescent cells, a quantity that is independent of the background fluorescence variation.

\begin{figure}[htb]
\begin{center}
\includegraphics[width= 0.99\textwidth]{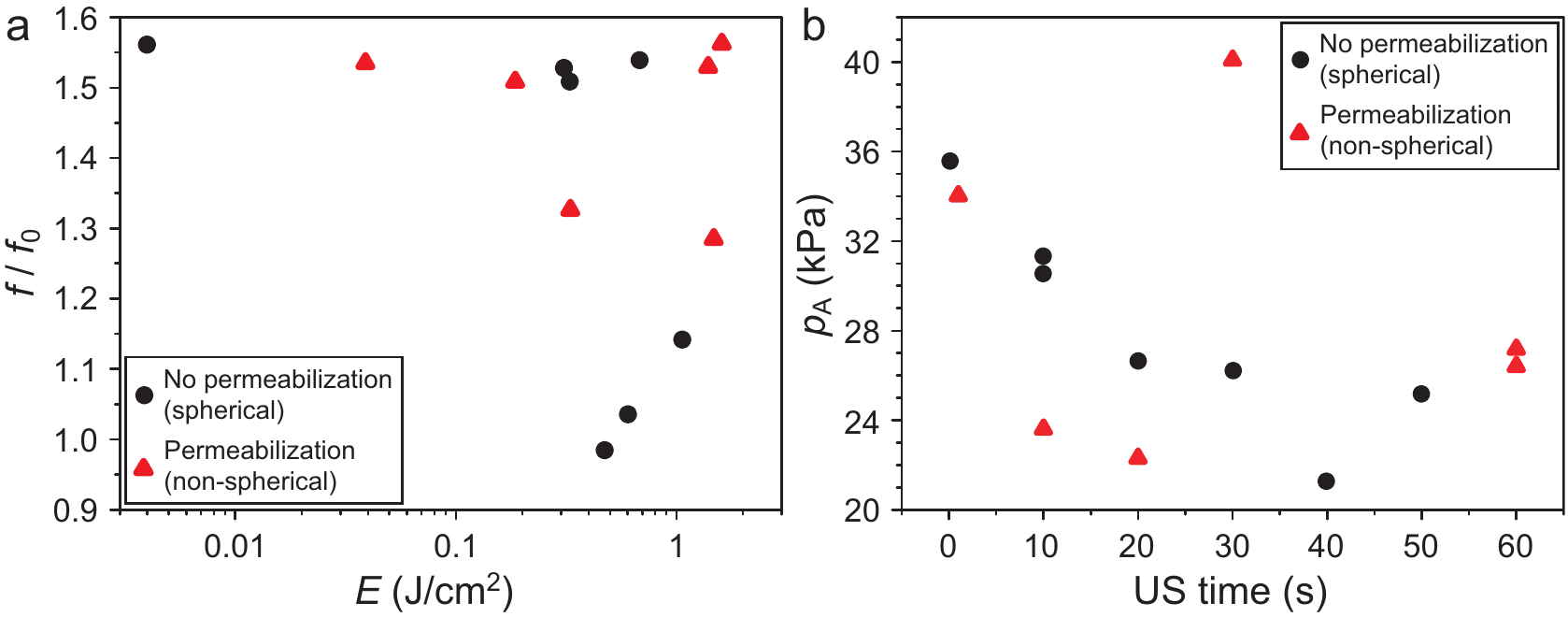}
\caption{Permeabilization (red triangles) and no permeabilization (black circles) events of algal cells under microstreaming flow as function of: a) the energy exposure and of the US frequency normalized by the resonance frequency $f_0$; b) the ultrasound exposure time and the acoustic pressure $p_A$. Black circles indicate events where the microstreaming flow was generated by spherical oscillations whereas red triangles indicate events characterized by microstreaming flow induced by non-spherical oscillations.} \label{fig:exposure}
\end{center}
\end{figure}

\begin{figure}[htb]
\begin{center}
\includegraphics[width= 0.99\textwidth]{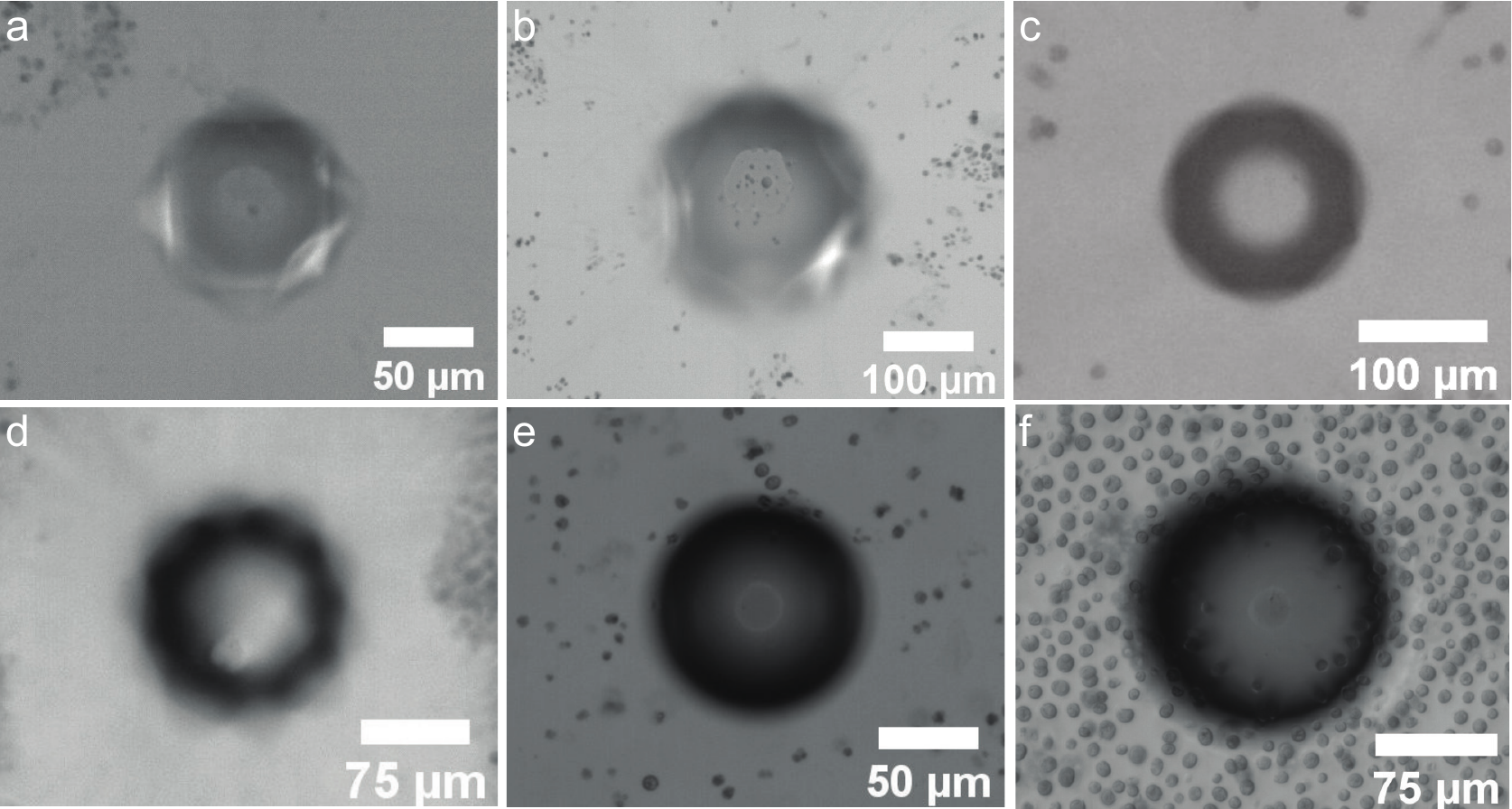}
\caption{a-d) Microbubbles under US field showing different surface modes $n$ during non-spherical oscillations. Acoustic pressure and frequency are: a) $p_A$ = 56.3 kPa and $f$ = 51 kHz; b) $p_A$ = 50.3 kPa and $f$ = 27 kHz; c) $p_A$ = 47.5 kPa and $f$ = 53 kHz; d) $p_A$ = 85.4 kPa and $f$ = 45 kHz; d, e) Microbubbles at rest showing algal cells attached (e) or not (f) to their surface before the microstreaming flow activation.} \label{fig:oscillations}
\end{center}
\end{figure}

In Figure~\ref{fig:exposure} we report the results of several permeabilization experiments on algal cells subjected to microstreaming flow induced by different microbubbles. In the  conditions of our experiments, the maximum rate of permeabilization that could be achieved was approximately 15\%. In Figure~\ref{fig:exposure}a the results are presented as a function of the acoustic energy exposure $E$ and of the applied ultrasound frequency, $f$, normalized by the resonance frequency $f_0$ of the microbubble. The acoustic energy exposure $E$ is obtained by multiplying the intensity $I$ of the acoustic wave (assuming a plane wave) by the ultrasound exposure time $t$ \cite{Azencott2007}:
\begin{equation}
E=I t=\frac{p_A^2}{2\rho_w c} t,
\label{eq:energy}
\end{equation}
with $\rho_w$ the water density and $c$ the speed of sound in water. We report our experimental data as a function of the acoustic energy exposure for direct comparison with bulk ultrasonication experiments of the same cells \cite{Azencott2007}. In Figure~\ref{fig:exposure}a we observe that cell wall permeabilization is obtained in a range of energy exposures between 0.04 and 1.5 J/cm$^2$. These values are between 10 and 100 times lower than those reported for bulk ultrasonication \cite{Azencott2007} where permeabilization of \textit{C. reinhardtii} at a rate of 10-20\% (measured by uptake of bovine serum albumin) was obtained for energy exposure in the range 25-75 J/cm$^2$. The main difference between the two experiments is that we used a pre-existing bubble, like in~\citet{Krehbiel2014}, and we used an ultrasound frequency close to the natural resonance of the bubble. Figure~\ref{fig:exposure}a shows also that the bubble oscillations can be either spherical or non-spherical, as confirmed by video microscopy for each experiment. Different predominant modes characterizing the shape oscillations could be recognized from the stroboscopic effect obtained acquiring images at 0.1 s. We observed several modes up to $n=10$ as shown in Figure~\ref{fig:oscillations}(a)-(d). In Figure~\ref{fig:exposure}a, spherical or non-spherical oscillations seem to occur without a clear dependence on frequency or acoustic energy exposure. Permeabilization events were observed only under microstreaming flow generated by microbubble dynamics characterized by non-spherical oscillations, whereas flows induced by microbubble spherical oscillations did not enhance algae permeabilization. The observation that the majority of permeabilization events occur with non-spherical oscillations is explained by the higher viscous stresses as compared to spherical oscillations, for the same values of frequency and acoustic pressure.

 The same results are presented in Figure~\ref{fig:exposure}(b) as a function of acoustic pressure amplitude, $p_A$, and ultrasound exposure time. All permeabilization events (triangles) were obtained for an acoustic pressure between 20 kPa and 40 kPa, almost two orders of magnitude lower than the pressure range used for bubble cavitation (1 MPa and higher). Algae permeabilization is induced by bubble dynamics characterized by non-spherical oscillations independently of the applied acoustic pressure and ultrasound exposure time. In these conditions we can obtain permeabilization within 20 seconds at pressures as low as 22.3 kPa (corresponding to $E=0.33$ J/cm$^2$) or within shorter times (1 s) by increasing the pressure to 34.0 kPa (corresponding to $E=0.04$ J/cm$^2$). Again no clear trend is found for the emergence of shape oscillations as a function of acoustic pressure.

Bright-field video microscopy provides additional insights in the mechanism of occurrence of shape oscillations in this system [Figure~\ref{fig:oscillations}(e-f)]. Shape oscillations of ultrasound-driven bubbles in a pure fluid are triggered as parametric instability, and driven by the bubble radial oscillations \cite{Versluis2010}. In our experiments we observe shape oscillations near the bubble resonance frequency, which is an indication that they are driven by the radial oscillations; we also found that it takes a finite time (from less than 1 second to tens of seconds) for the shape oscillations to fully develop, which is also consistent with a parametric instability. Having confirmed the nature of the observed shape oscillations as parametric instability, we can explain why we observed their development only in some microbubbles with no clear dependence on pressure and frequency (see Figure~\ref{fig:exposure}). A parametric instability initially manifests itself as a small perturbation of the spherical interface, which then grows with time. We identified that, in our system, the origin of these small perturbations is the presence of algal cells on the surface of the bubble prior to the activation of the microstreaming flow, as shown in Figure~\ref{fig:oscillations}(e). All microbubbles showing algae on their surface developed non-spherical shape oscillations, whereas those without algae on the surface (see Figure~\ref{fig:oscillations}(f)) did not. Other microorganisms have also been observed to attach to fluid interfaces due to surface tension forces \cite{Stebe2020}. It also appears that presence of algal cells in the vicinity of the microbubble is not sufficient to develop surface modes; the cells must be on the bubble surface to trigger the instability.

The mechanism for enhanced cell wall permeability under the action of microbubble streaming flow remains to be elucidated; in our experiments we could not visualize pore formation. The rate of permeabilization was limited to $\sim$15\% by the range of experimental parameters accessible with our experimental setup. Nevertheless, it is remarkable that we obtained cell wall permeabilization for a mechanical energy input 10-100 times lower than previously reported for bulk ultrasonic treatment. This finding opens the way to energy-efficient cell wall permeabilization for bioproduct recovery and holds potential for new analytical tools for manipulation of microorganisms of use to a variety of domains. 
\section*{Conflict of Interest Statement}
The authors have no conflicts to disclose.

\begin{acknowledgments}
We thank A. Jamburidze and A. Pouliopoulos for assistance with the experiments. This work was supported in part by the Seventh Framework Programme of the European Commission (Grant Agreement no. 618333).  
\end{acknowledgments}

\section*{Data Availability Statement}
The data that support the findings of this study are available from the corresponding author upon reasonable request.

\bibliography{bibliography}

\section*{SUPPORTING INFORMATION}

\subsection*{a. Algal cell cultures}
Stock cultures of wild-type \textit{C. reinhardtii} strain cc-124 were obtained on solid agar. Fresh algae suspensions were prepared by resuspending cells from the stock culture in 25~mL of sterile medium at room temperature, under cool white fluorescent light, and with a starting pH~=~7. The culture was mechanically agitated to provide aeration. After 5 days of growth, the culture was scaled up to a volume of 250~mL of fresh medium (10\%~v/v inoculum). The entire growth process lasted two weeks. The culture was then scaled up to a volume of 2-3~L and transferred in a vertical photobioreactor. The concentration of algal cells was determined from optical density measurements at 663~nm using a UV-vis spectrophotometer. Once the cell concentration reached a value of approximately 0.3~mg~ml$^{-1}$, the algae were grown under nitrogen starvation conditions for one week to enhance the lipid content. Algal cells suspensions with a final concentration between 0.3 and 0.8~mg~ml$^{-1}$ were used in the experiments. Sterile conditions were maintained at all stages to ensure the purity of the strains. Contamination was monitored routinely by both microscopy and spread plate technique.
\subsection*{b. Fluorescent staining of cells}
BODIPY (Difluoro {2-[1-(3,5-dimethyl-2H-pyrrol-2-ylidene-N)ethyl]-3,5-dimethyl-1H-pyrrolato-N} boron) was used to stain intracellular lipid bodies. This dye was selected because its green emission peak does not interfere with the red autofluorescence of algal chloroplasts, and because its high oil/water partition coefficient promotes its transport across the thick algal cell wall \cite{Rumin2015}. A disadvantage of BODIPY is that it is not sensitive to the environment polarity and, being fluorescent also in water, the free dye in the aqueous cell suspension produces a background signal, which however does not affect our image analysis method (see below). BODIPY and dimethyl sulfoxide (DMSO) were purchased from Sigma Aldrich and used as received. Staining of the intracellular lipid bodies of \textit{C. reinhardtii} was performed according to the method described by \citet{Velmurugan2013}. Briefly, a 5~mM BODIPY stock was prepared by dissolving in DMSO and stored in the dark. An aliquot of BODIPY stock solution was added to the algal suspension, to give a final BODIPY concentration of 10~$\mu$M. The suspension was agitated for 1 minute with a vortex mixer. Samples were incubated in darkness conditions for 5~min at room temperature and immediately transferred to the sample cell for the experiments. The uptake of BODIPY in the absence of flow was quantified at the end of the 5-min incubation period, to determine a threshold above which further dye uptake can be ascribed to an effect of the applied flow. The ratio of number of stained cells to total number of cells was found to be always lower than 5\%. Therefore, the threshold value for cell wall permeabilization due to the acoustic microstreaming flow was set to 5\%. 
\subsection*{c. Image analysis}
The images before and after the microstreaming flow were analysed using the open-source software ImageJ. The analysis consisted in counting the number of fluorescent cells in the field of view before and after the flow, to determine the percentage of fluorescent cells over a total number of 1000 cells on average per image. The cell counting was done by selecting in every image a threshold value in the gray scale and considering only the objects characterised by gray values aboce this threshold. The threshold value was selected according to the histogram of the image as shown in Figure~\ref{fig:SI1} where the histograms of images shown in Figure~\ref{fig:fluorescence}(a) and (b) are reported. As indicated by the red dashed line, we set the threshold between the peak with lower gray value and high counts ($>$ 10$^4$), corresponding to the background fluorescence, and the smaller peak (counts $<$ 10$^3$, see insets of Figure~\ref{fig:SI1}) at higher gray values, coming from the fluorescent algal cells. In this way we generated a mask with only the fluorescent cells for each image in order to count their number. This counting protocol allowed us to discard the background fluorescence from all images in a consistent way, thus avoiding that the background fluorescence variation could affect the image analysis.

The field of view for cell counting includes an area around the microbubble at least one bubble diameter away from the surface of the bubble, to visualize the entire area subjected to the microstreaming flow. We limited the counting to algae that sedimented on the bottom because, unlike free swimming algal cells, they remained in the field of view during the entire experiment duration ($\leq$ 60 s). Moreover, in order to allow the algal cells entrained in the flow streamlines to sediment (see Figure~\ref{fig:method}(e)), we waited 1 minute after stopping the flow and before counting. Inevitably, also other cells not affected by the flow will sediment during this waiting time. Their counting led to an underestimation of the effective cell permeabilization induced by the microstreaming flow. Nevertheless, we could successfully prove algal cell wall permeabilization.

\begin{center}
\includegraphics[width= 0.99\textwidth]{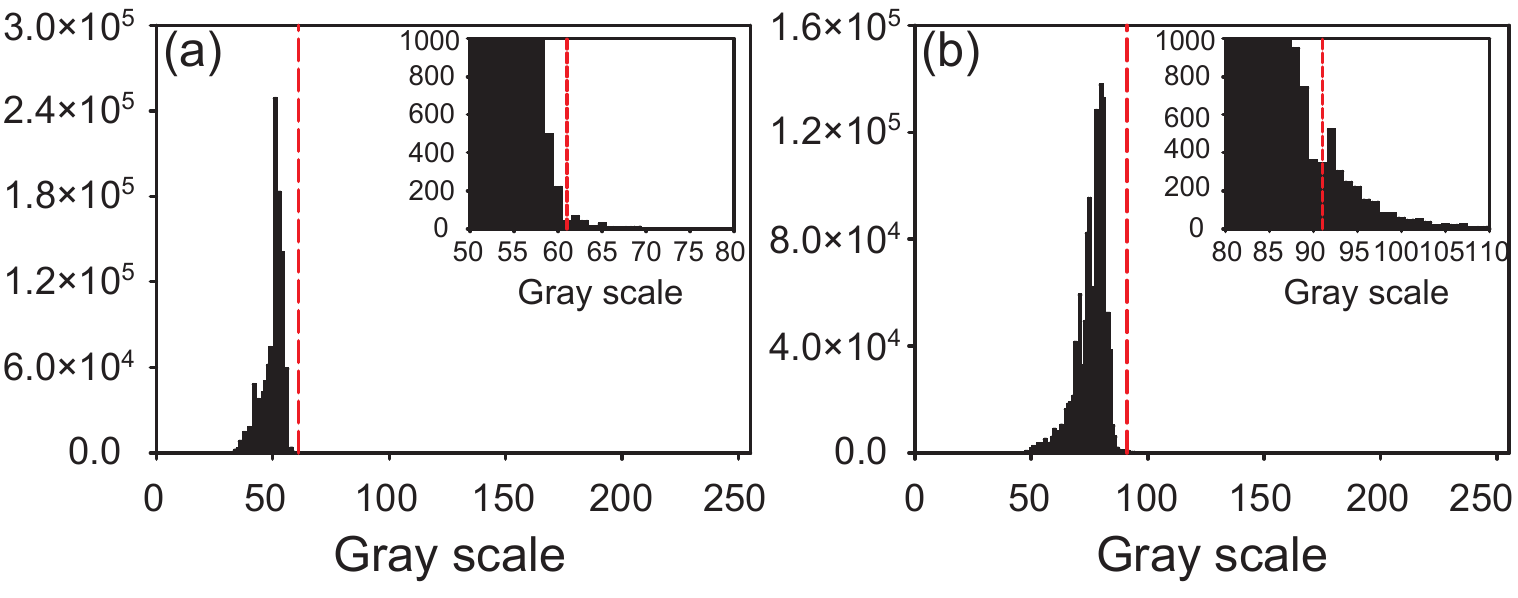}
\end{center}
\small{\textbf{SI Figure 1.} Histograms of the images of Figure2a and 2b. The red dashed lines show the threshold gray values used to make the masks shown in Figure2c and 2d. The two insets highlight the presence of the peaks coming from the fluorescent algal cells.} \label{fig:SI1}

\end{document}